\documentclass[twocolumn,showpacs,preprintnumbers,amsmath,amssymb]{revtex4}

\usepackage{graphicx}
\usepackage{dcolumn}
\usepackage{bm}

\begin{document}

\preprint{IUHET-501}

\title{Quantum Mechanics and the Generalized Uncertainty Principle}

\author{Jang Young Bang}
 \email{jybang@indiana.edu}
\author{Micheal S. Berger}%
 \email{berger@indiana.edu}
\affiliation{%
Physics Department, Indiana University, Bloomington, IN 47405, USA}

\date{\today}

\begin{abstract}
The generalized uncertainty principle has been described as a general 
consequence of incorporating a minimal length from a theory of quantum 
gravity. We consider a simple quantum mechanical model where the operator
corresponding to position has discrete eigenvalues and show how the 
generalized uncertainty principle results for minimum uncertainty wave 
packets.
\end{abstract}

\maketitle

\def\al{\alpha}
\def\be{\beta}
\def\ga{\gamma}
\def\de{\delta}
\def\ep{\epsilon}
\def\ve{\varepsilon}
\def\ze{\zeta}
\def\et{\eta}
\def\th{\theta}
\def\vt{\vartheta}
\def\io{\iota}
\def\ka{\kappa}
\def\la{\lambda}
\def\vpi{\varpi}
\def\rh{\rho}
\def\vr{\varrho}
\def\si{\sigma}
\def\vs{\varsigma}
\def\ta{\tau}
\def\up{\upsilon}
\def\ph{\phi}
\def\vp{\varphi}
\def\ch{\chi}
\def\ps{\psi}
\def\om{\omega}
\def\Ga{\Gamma}
\def\De{\Delta}
\def\Th{\Theta}
\def\La{\Lambda}
\def\Si{\Sigma}
\def\Up{\Upsilon}
\def\Ph{\Phi}
\def\Ps{\Psi}
\def\Om{\Omega}
\def\mn{{\mu\nu}}
\def\cD{{\cal D}}
\def\cF{{\cal F}}
\def\cL{{\cal L}}
\def\cS{{\cal S}}
\def\fr#1#2{{{#1} \over {#2}}}
\def\frac#1#2{\textstyle{{{#1} \over {#2}}}}
\def\pt#1{\phantom{#1}}
\def\prt{\partial}
\def\vev#1{\langle {#1}\rangle}
\def\ket#1{|{#1}\rangle}
\def\bra#1{\langle{#1}|}
\def\amp#1#2{\langle {#1}|{#2} \rangle}
\def\half{{\textstyle{1\over 2}}}
\def\lsim{\mathrel{\rlap{\lower4pt\hbox{\hskip1pt$\sim$}}
    \raise1pt\hbox{$<$}}}
\def\gsim{\mathrel{\rlap{\lower4pt\hbox{\hskip1pt$\sim$}}
    \raise1pt\hbox{$>$}}}
\def\ol#1{\overline{#1}}
\def\Re{\hbox{Re}\,}
\def\Im{\hbox{Im}\,}
\def\etal {{\it et al.}}
\def\slash#1{\not\hbox{\hskip -2pt}{#1}}

\section{Introduction and Motivation}

There is reason to believe in the existence of
a minimal length that can in principle be measured. This viewpoint
stems in part from the realization that if enough mass-energy is confined to a 
small region of space, a black hole must
form. For example if one increases the energy of colliding particles 
beyond the Planck energy, one expects the short distance effects to be 
hidden behind an event horizon. In fact, as the energy is 
increased, the size of this event horizon increases. One way an
effective minimal length might arise is through the discretization of
spacetime. One might also ask the question whether the generic appearance of 
a minimal length in a low-energy effective theory of quantum gravity 
survives in the full theory when its ultraviolet sector is completed.

In a theory of quantum gravity a fundamental distance scale 
is expected to be of order the Planck length $L_P$.  
The existence of a minimal length invites the possibility that there are 
corrections to the usual Heisenberg uncertainty principle such that it 
becomes what is known as a generalized uncertainty principle (GUP):
\begin{eqnarray}
\Delta x \ge {{1}\over {2\Delta p}}+ \alpha L_P^2 \Delta p + \dots
\;.
\label{gup}
\end{eqnarray}
We have set $\hbar =1$.
The terms in the ellipsis represent higher order contributions which should 
generically be present.
They may involve higher order powers of $\Delta p$, but might also involve
expectation values of higher powers of the momentum.
We shall sometimes refer to the expression in Eq.~(\ref{gup}) minus these 
extra terms as the 
truncated GUP. The minimal length in the theory is governed by the 
parameter $\alpha$, and the generalized uncertainty principle in 
Eq.~(\ref{gup}) shows that there is a minimum dispersion $\Delta x$ for 
any value of $\Delta p$ at least as long as the first two terms on the 
right hand side are considered. 

Interest in a minimum length or the generalized uncertainty principle 
has been motivated by studies of the 
short distance behavior of 
strings\cite{Veneziano,Gross:1987ar,Amati:1988tn,Konishi:1989wk}, 
considerations regarding the properties of black holes\cite{Maggiore:1993rv},
and de Sitter space\cite{Snyder:1946qz}.
In recent years the
generalized uncertainty principle has been studied extensively in the 
literature\cite{Silvert:1970kr,Adler:2001vs,Medved:2004yu,Nozari:2005it,Chang:2001kn,Kempf:2000ac}. 
Most of the research does not attempt to derive the uncertainty principle from
quantum 
gravity explicitly. Rather the modifications to the uncertainty relation
are motivated by what is a general property on any theory of quantum 
gravity, and the implications
of the GUP are analyzed phenomenologically. Other approaches have involved
trying to understand the quantum mechanical basis for the uncertainty principle
in Eq.~(\ref{gup}) often by considering a truncation of the full series of 
terms on the right hand side. Commutation relations can be postulated which 
give rise to this truncated series and the associated algebra has been studied.
In this paper we try to 
understand how the GUP can arise in a simple quantum mechanical model (which
is well-understood), and how the GUP is in fact completed when all the 
relevant terms are included. This means we derive all the terms in the 
ellipsis in Eq.~(\ref{gup}). The model contains an operator with discrete 
eigenvalues in the familiar way: because the phase space is compactified.

The model is a quantum theory with a position operator that has discrete 
eigenvalues, so it may be considered a theory with a minimum length associated
with the difference between eigenvalues. However we do not a priori exclude the
possibility that a physical state can be an eigenstate of this operator.
That leads us to consider quantum mechanics on a circle. 
Usually one is concerned with the 
problem of a spatial direction compactified such that the eigenvalues of the 
momentum operator are quantized. Here we (eventually)
consider the somewhat more 
unusual situation where
one considers the {\it momentum} to be compactified, thus resulting in a 
quantum mechanics with discrete eigenvalues for the operator associated with 
a spatial coordinate. 
Quantum mechanics on a circle is a simple physical problem which has a long 
history, and does not contain gravity 
explicitly. The 
compactification of momentum does provide a discrete spectrum for the position
coordinate which may be a feature of a full theory of quantum gravity. 
We are able to derive 
the generalized uncertainty relation in Eq.~(\ref{gup}) and calculate the 
coefficient $\alpha L_P^2$ in terms of the compactification ``radius'' 
of the momentum circle. In fact we are able to obtain straightforwardly 
the whole series of
higher order terms (i.e. higher powers of $\Delta p$ and factors like
$\langle p^n\rangle $) on the right hand side of the
relation for minimum uncertainty wave packets. 
Since
the terms explicitly expressed in Eq.~(\ref{gup}) should represent only an 
approximation to a full expression, this may shed some light (at least as far
as the simple model captures the necessary features of a full theory of 
gravity) on how the GUP fits into a complete theory of discretized 
spacetime. 
We believe this derivation of a generalized uncertainty 
principle in an elementary context may be useful in understanding what 
features of a discrete spacetime may be needed to obtain the uncertainty 
principle in a more realistic context. Quantum mechanics on a  
circle has been chosen simply for 
simplicity and because it is well-studied in other contexts.
It makes the presentation particularly elementary since 
modified Bessel functions are needed to describe the wavefunctions and the 
coefficients of the eigenfunctions of the minimum uncertainty wave packets.

The study of the proper way to treat quantum mechanics on a compact
space goes back to the earliest days of the subject\cite{London,Jordan}.
A history of the subject can be found in Ref.~\cite{Kastrup:2003fs}.
Therefore it has been known
for a long time that the uncertainty principle as derived for a 
circular coordinate (phase) operator and the angular momentum operator
is inadequate because the angular phase variable is periodic.
More recent attempts
to rectify this problem led to various definitions for the phase 
dispersion. Judge defined\cite{Judge}
\begin{eqnarray}
\Delta \theta^2 = \min _{-\pi\le\gamma\le\pi} 
\int^\pi_{-\pi}\theta^2|\psi(\gamma +\theta)|^2d\theta \;,
\end{eqnarray}
for which he conjectured the uncertainty relation
\begin{eqnarray}
\Delta L \cdot {{\Delta \theta}\over {1-(3/\pi^2)\Delta \theta^2}} \ge
{1 \over 2}\;.
\label{judge}
\end{eqnarray} 
This conjecture was later proved by Bouten, Maene and Van Leuven\cite{Bouten},
and discussed further in Ref.~\cite{Chisolm}.
The form of the uncertainty relation in Eq.~(\ref{judge}) 
allows for an angular momentum eigenstate (for which $\Delta L=0$) to 
satisfy the inequality because the bounded uncertainty in $\Delta \theta^2$
causes the denominator to vanish for these eigenstates.
The uncertainty relation in Eq.~(\ref{judge}) allows for a treatment of the 
case of an angular momentum eigenstate for which $\Delta L=0$ whereas the 
dispersion in the angle is bounded $\Delta \theta =\pi^2/3$.

The subtle issue of an uncertainty principle involving the phase 
operator was subsequently discussed in a classic paper, Ref.~\cite{Susskind}. 
There, Hermitian cosine and sine operators were
defined and various uncertainty principles were suggested. 
The crucial realization is that the angular coordinate variable $\theta$ is 
not suitable for quantization, and one should use a phase operator 
$e^{i\theta}$ (or alternatively the cosine and sine operators).
This early important work is developed and 
reviewed in Ref.~\cite{Carruthers:1968my}.
This has been discussed and reviewed further and applications to more general
situations have appeared in 
Refs.~\cite{Jackiw,Levy-Leblond:1976uj,Newton:1979ut}.
In the formulation of Ohnuki and 
Kitakado\cite{Ohnuki:1991uv,Tanimura:1993hf} 
it was shown that there are in fact an infinite number of 
representations of the algebra of operators which can be understood in 
terms of a certain gauge field. These representations are classified by 
the value of a parameter $\alpha \in [0,1)$ which specifies the gauge 
inequivalent representations and interpolates between the discrete 
eigenvalues in the operator spectrum. 
Finally in Ref.~\cite{Tanimura:1993hf} the minimum uncertainty
wave packets on the circle
were shown to reduce in the large radius limit to the usual Gaussian 
wave packet. It is this limit of a large radius that will most concern 
us here, as we will show how the usual Heisenberg uncertainty principle
expressed in terms of $\Delta x$ and $\Delta p$ on the line get modified
in the large radius limit in which the corrections should be 
small. One potential benefit of identifying the generalized uncertainty 
principle in this limit is that one can extend the results to cases that 
are not approximately quantum mechanics on the line, and study how the 
minimum uncertainty wave packets
on the circle interpolate between the Gaussian wave packet 
on the line to the (angular) momentum eigenstate on the circle. This extends
the generalized uncertainty principle in Eq.~(\ref{gup}), at least for this 
simple case, to physical situations that do not require the ``extra'' terms to 
be subleading in magnitude.

Finally we note that the minimum uncertainty wave packets 
we examine are well-known in the field
of quantum optics where they are sometimes called (circular) squeezed states. 
This is a 
rich and well-developed subject and we refer the reader to 
Ref.~\cite{Kastrup:2003fs} for an overview. Coherent and squeezed
states have been 
developed for general Lie symmetries. Important examples are 
the Barut-Girardello
coherent states\cite{Barut:1970qf}.
See also 
Refs.~\cite{Bluhm:1995nr,Trifonov:2000su,Kastrup:2001hx,Gonzalez:1998kj,Kowalski:1998hx,Roy:1982nu,Santhanam:1976zs}. Our emphasis is different, however, 
in that we are primarily 
concerned with the properties of the uncertainty relations for a 
compactified phase space in the large radius
limit. It is in this limit that the corrections to the usual 
Heisenberg uncertainty relation involving a position operator $x$ and
a momentum operator $p$ can be understood to be small.
In a more general context one is interested in how to quantize a system with 
some classical phase space. 
Of particular importance in understanding the correct algebra for the quantum
system is the group theoretic quantization described in detail in 
Ref.~\cite{Isham}. Quantum mechanics on a circle serves as a simple example of 
this geometric quantization.

This paper is organized as follows. In Section II we review the quantum 
mechanics formulated on a circle as formulated by Ohnuki and Tanimura
emphasizing the features of importance to us. In Section III we derive a
generalization of the Heisenberg uncertainty involving $\Delta x$ and 
$\Delta p$ for the case where the space is compactified on a circle and 
the eigenvalues of the momentum operator are discrete. 
We show how all the terms in an infinite series are known. 
In Section IV we imagine that the momentum is compactified, so that the 
roles of the position and momentum operators are interchanged. In this 
case the usual form of the GUP results. In Section V we give some explicit 
numerical results for 
the minimum uncertainty wave packet and briefly examine for which states 
the generalized uncertainty principle as expanded in Eq.~(\ref{gup}) is 
a reasonable expansion with decreasing terms. We summarize in Section V.

\section{Quantum Mechanics on the Circle}

We first review quantum mechanics on a circle. We use
the particular  formulation given 
by Ohnuki and Kitakado\cite{Ohnuki:1991uv} and explored further in 
Ref.~\cite{Tanimura:1993hf}. There is a Hermitian operator $G$ and a 
unitary operator $W$ satisfying the commutation relation 
\begin{eqnarray}
\left [ G, W\right ]=W.
\label{comm}
\end{eqnarray}
The operators $G$, $W$ and $W^\dagger$ form the fundamental algebra of 
quantum mechanics on the circle. The algebra indicates that the operators
$W$ and $W^\dagger$ act as raising and lowering operators on the eigenstates
of $G$. If $|\alpha \rangle$ is an eigenstate of the operator $G$, 
\begin{eqnarray}
G|\alpha \rangle =\alpha |\alpha \rangle\;, 
\end{eqnarray}
then
\begin{eqnarray}
GW|\alpha \rangle &=&(\alpha +1)W|\alpha\rangle \nonumber \\
GW^\dagger |\alpha \rangle &=&(\alpha -1)W^\dagger |\alpha\rangle  
\end{eqnarray}
The operator $W$ can be called the phase 
operator because it has the eigenvalue solution
\begin{eqnarray}
W|\theta \rangle = e^{i\theta}|\theta\rangle\;.
\end{eqnarray}
The solution is 
\begin{eqnarray}
|\theta \rangle =\kappa (\theta)\sum _{n=-\infty}^{+\infty}e^{-in\theta}
|n+\alpha \rangle\;,
\end{eqnarray}
where $\kappa(\theta)$ is a periodic phase, i.e. $|\kappa (\theta)|=1$ and
$\kappa(\theta +2\pi)=\kappa (\theta)$. It can be 
shown\cite{Ohnuki:1991uv,Tanimura:1993hf}
that one may choose $\kappa(\theta)=1$ which is a kind of gauge choice.
Finally there is the representation of action of the operator $W$ on a 
wave function,
\begin{eqnarray}
\langle \theta | W | \psi \rangle = e^{i\theta}\psi(\theta)\;.
\end{eqnarray}

One can form the Susskind-Glogower operators
\begin{eqnarray}
C&=&{1\over 2}\left (W+W^\dagger \right )\;, \nonumber \\
S&=&{1\over {2i}}\left (W-W^\dagger \right )
\end{eqnarray}
which are Hermitian. Carruthers and Nieto\cite{Carruthers:1968my} studied 
uncertainty relations involving the operators 
$C$ and $S$ in a classic paper. Here we do not try to devise 
uncertainty relations involving $G$, $W$ and $W^\dagger$ (or equivalently $G$, 
$C$ and $S$), but rather make an identification between quantum mechanics on 
the circle when the large radius limit is taken and quantum 
mechanics on the line.

Applying the Schwarz inequality to the states $\Delta G|\psi\rangle$ and 
$\Delta W|\psi\rangle$ one obtains the uncertainty relation
\begin{eqnarray}
\langle \Delta G^2\rangle \langle \Delta W^\dagger \Delta W\rangle 
\ge |\langle \Delta G \Delta W\rangle |^2\;.
\label{unc0}
\end{eqnarray}

\begin{figure}[t]
\centerline{
\mbox{\includegraphics[width=3.50in]{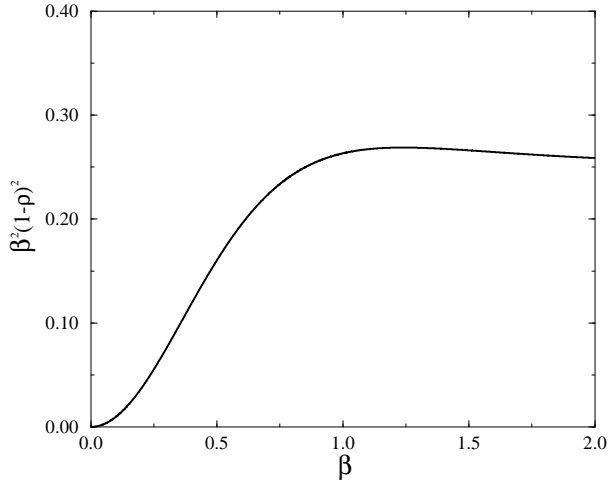}}
}
\caption{The right hand side of the 
uncertainty relation as a function of $\beta$. It approaches $1/4$ as 
$\beta\to\infty$ as expected, and goes to $0$ for $\beta=0$ which corresponds 
to an angular momentum eigenstate.}
\label{rhs}
\end{figure}

The minimum uncertainty wave packet expressed in the 
notation of Ref.~\cite{Tanimura:1993hf} is
\begin{eqnarray}
\psi(\theta)={1\over {\sqrt{I_0(2\beta)}}}\exp \left [\beta e^{i(\theta -\phi)}
+i\nu \theta \right ]\;.
\label{muwp}
\end{eqnarray}
This state saturates the uncertainty relation, Eq.~(\ref{unc0}), and is a 
state peaked at $\theta=\phi$.
The normalization convention adopted here is 
\begin{eqnarray}
\langle \psi |\psi\rangle ={1\over {I_0(2\beta)}}\int_0^{2\pi}
{{d\theta}\over {2\pi}}e^{2\beta\cos(\theta -\phi)}=1\;.
\end{eqnarray}
The paramter $\nu$ must take on an integer value so that $\psi(\theta)$ is
single-valued. 
The variances are 
\begin{eqnarray}
\langle \Delta W \Delta W^\dagger \rangle &=&1-\rho^2\;, \nonumber \\
\langle \Delta G^2 \rangle &=&\beta^2(1-\rho^2)\;,
\end{eqnarray}
so that the uncertainty relation is expressed as 
\begin{eqnarray}
\langle \Delta G^2 \rangle \langle \Delta W \Delta W^\dagger \rangle =
\beta^2(1-\rho^2)^2\;,
\label{unc}
\end{eqnarray}
where 
\begin{eqnarray}
\rho={{I_1(2\beta)}\over {I_0(2\beta)}}\;,
\end{eqnarray}
and $I_n(z)$ are the modified Bessel functions. 
The right hand side of this relation is plotted in Fig.~\ref{rhs}. 
The notable 
features of this expression are that in the $\beta\to \infty$ ($\rho\to 1$)
limit, it 
approaches $1/4$ as expected in the large radius limit. In the other limit
$\beta\to 0$ ($\rho\to 0$), 
it goes to zero as shown in Fig.~\ref{rhs}. 
This is the well known case where the 
wave function is an eigenstate of (angular) momentum, so that there is zero 
dispersion. One also has 
\begin{eqnarray}
\beta^2(1-\rho^2)^2={1\over 4}+{\cal O}\left ({1\over \beta^2}\right )\;,
\label{expand}
\end{eqnarray}
in the large $\beta$ limit. This indicates that in the large $\beta$ limit
the variances for the operators $G$ and $W$ can be interpreted in terms of
the usual operators on the line, namely $x$ and $p$, and that the 
Heisenberg uncertainty relation involving $x$ and $p$ can be obtained in 
that limit.

On the circle the relevant operator is the phase operator $W$, 
and one encounters the well-known problems (briefly described in the 
introduction) if one insists on using the 
angle as a operator. However
in the large radius limit one can 
express an effective relationship between the dispersion in $\theta$ and the 
dispersion in momentum. In this limit the GUP appears as 
the first terms in an infinite expansion.

The expectation values and the variances for the operators in question were
derived in Ref.~\cite{Tanimura:1993hf}. The expectaion values are
\begin{eqnarray}
\langle W \rangle &=&\rho e^{i\phi}\;, \nonumber \\
\langle G \rangle &=&\nu +\alpha +\beta \rho\;.
\end{eqnarray}
The parameter 
$0\le \alpha < 1$ labels the inequivalent representations of the quantum 
mechanics algebra in Eq.~(\ref{comm}).
It can be viewed as an interpolation between the discrete eigenvalues of the 
operator $G$.

If one defines the variables $x$, $\langle x \rangle$ and $\langle p \rangle$
as 
\begin{eqnarray}
x&=&r\theta\;,\nonumber \\
\langle x \rangle &=& r\phi\;, \\
\langle p \rangle &=& {1\over r}\langle G \rangle\;, \nonumber 
\label{defx}
\end{eqnarray}
then in the large radius limit one recovers the (minimum uncertainty)
Gaussian wave packet\cite{Tanimura:1993hf} 
\begin{eqnarray}
\Psi(x)=\left ({1\over {2\pi d^2}}\right )^{1/4}\exp
\left [-{1\over {4d^2}}(x-\langle x\rangle)^2+i\langle p \rangle
(x-\langle x\rangle)\right ]\;,
\end{eqnarray}
when one makes the identification
\begin{eqnarray}
{\beta\over r^2}={1\over {2d^2}}\;,
\label{defd}
\end{eqnarray}
and defines the normalization condition
\begin{eqnarray}
|\psi(\theta)|^2{{d\theta}\over {2\pi}}=|\Psi(x)|^2dx\;.
\end{eqnarray}
One can calculate the first order corrections (in $1/\beta$ or 
$1/r^2$) to find the modifications the finite size of the circle give to the 
Gaussian packet. Given a compactification radius $r$, then the degree the 
minimum uncertainty wave
packet is localized is determined by either the parameter $\beta$ or 
the parameter $d=r/\sqrt{2\beta}$ according to Eq.~(\ref{defd}).
These modifications in fact are of the form suggested by 
the GUP.

The momentum eigenstates are quantized so that the consecutive eigenvalues of 
the operator $G$ differ by unity. So the momentum eignevalues are separated by 
$1/r$ according to Eq.~(\ref{defx}).

\section{Derivation of a Generalized Uncertainty Principle}

One expects on general grounds in any theory of quantum gravity 
that a GUP may apply when the momenta are of order the Planck scale and 
gravitational effects become important. In this section
we demonstrate how in a simple quantum mechanical model with discrete space
eigenvalues the GUP can be simply derived. Since our model does not contain
gravity explicitly, the GUP will be seen to arise as a consequence of the 
discretization of space which may or may not be a property of quantum 
gravity. Since the toy model we utilize can be solved, the 
completion of the GUP in the ultraviolet limit can be derived and in fact
the entire expansion of higher order terms in the GUP can be explicitly 
derived. This ``derivation'' of the GUP in a simple model may be useful in 
understanding how the GUP arises in more realistic physical situations.

The dispersion that we are interested in involves a parameter $x=r\theta$ 
where $r$ is the compactification radius. The coordinate $x$ is periodic 
as is $\theta$, but for sufficiently localized wave packets ($d<<r$),
it can be used as an effective coordinate.
The uncertainty principle for the operators $W$ and $G$ can be expressed in 
terms of a series expansion involving the expectation values of the
angular parameter $\theta$. 
For a sufficiently localized wave packet ($d<<r$ or $\beta >>1$) the expansion
will involve increasing smaller terms.
We have 
\begin{eqnarray}
\langle W \rangle &=&\int_0^{2\pi}
{{d\theta}\over {2\pi}} \psi^* \psi e^{i\theta}\nonumber \\
&=&\int_0^{2\pi}{{d\theta}\over {2\pi}} 
\psi^* \psi(1+i\theta-{\theta^2\over 2}-i{\theta^3\over 6}+\dots)
\nonumber \\
&=&1+i\langle \theta \rangle -{{\langle \theta ^2 \rangle}\over 2}
-i{{\langle \theta ^3 \rangle}\over 6}+\dots
\end{eqnarray}
and 
\begin{eqnarray}
\langle W^\dagger \rangle &=&\int_0^{2\pi}
{{d\theta}\over {2\pi}} \psi^* \psi e^{-i\theta}\nonumber \\
&=&\int_0^{2\pi}{{d\theta}\over {2\pi}}
\psi^* \psi(1-i\theta-{\theta^2\over 2}+i{\theta^3\over 6}+\dots)
\nonumber \\
&=&1-i\langle \theta \rangle -{{\langle \theta ^2 \rangle}\over 2}
+i{{\langle \theta ^3 \rangle}\over 6}+\dots
\end{eqnarray}
For a localized wave packet, these series contain terms that are 
increasingly smaller\footnote{Note that $\langle \theta^2 \rangle$ coincides 
with the definition of Judge\cite{Judge} if coordinates are chosen so that 
$\langle \theta \rangle =0$.}. 
Then
\begin{eqnarray}
1-\langle W\rangle \langle W^\dagger \rangle &=&\langle \theta ^2\rangle
-\langle \theta \rangle ^2+\dots \nonumber \\
&=&{1\over r^2}\left (\langle x^2\rangle -\langle x \rangle ^2\right )+\dots
\nonumber \\
&=&{1\over r^2}\Delta x^2+\dots \;,
\end{eqnarray}
where the omitted terms involve terms which are smaller such as 
$\langle \theta^4\rangle$ and  $\langle \theta^2 \rangle^2$.
In fact we calculate
\begin{eqnarray}
1-\langle W\rangle \langle W^\dagger \rangle &=&
{1\over r^2}\left (\langle x^2\rangle -\langle x \rangle ^2\right )
\nonumber \\
&&+{1\over r^4}\left (-{1\over {12}}\langle x^4\rangle 
+{1\over 3}\langle x^3 \rangle \langle x \rangle
-{1\over 4}\langle x^2 \rangle ^2\right )
\nonumber \\
&&+{\cal O}\left ({1\over {r^6}}\right )\;.
\label{Wunc}
\end{eqnarray}
Using also $\langle \Delta p^2 \rangle={1\over r^2}\Delta G^2$ 
we get an uncertainty relation from Eq.~(\ref{unc0}),
\begin{eqnarray}
\Delta p^2 &&\Bigg [\left (\langle x^2\rangle -\langle x \rangle ^2\right )
\nonumber \\
&&+{1\over r^2}\left (-{1\over {12}}\langle x^4\rangle 
+{1\over 3}\langle x^3 \rangle \langle x \rangle
-{1\over 4}\langle x^2 \rangle ^2\right )
\nonumber \\
&&+{\cal O}\left ({1\over {r^4}}\right )\Bigg ]=\beta^2(1-\rho^2)^2\;.
\label{px}
\end{eqnarray}
We can without loss of generality choose our coordinate system so that 
$\langle x \rangle =0$. This is equivalent to taking $\phi=0$ in 
Eq.~(\ref{muwp}), and centers the wave packet at $x=0$, so that 
$\Delta x^2=\langle x^2 \rangle$. One then obtains 
\begin{eqnarray}
\Delta p^2 \Delta x^2 \Bigg [1 
-{1\over {4r^2}}(\Delta x^2)&-&{1\over {12r^2}}{{\langle x^4\rangle}\over 
{\Delta x^2}}
+\dots \Bigg ]=
\nonumber \\
&&\beta^2(1-\rho^2)^2\;,
\end{eqnarray}
where the ratio ${\langle x^4\rangle}/{\Delta x^2}$ is a calculable constant.
For the approximation we are contemplating ($\Delta x^2/r^2<<1$) one gets
\begin{eqnarray}
\Delta p^2 \Delta x^2 ={1\over 4}\left (1+{1\over {4r^2}}(\Delta x^2)
+{1\over {12r^2}}{{\langle x^4\rangle}\over 
{\Delta x^2}}
+\dots\right )
\;,
\end{eqnarray}
where we have made use of Eq.~(\ref{expand}).
It should be noted that the complete generalization of the uncertainty 
principle involves higher expectation values such as $\langle x^4\rangle$, 
and that the GUP is only an approximation 
to this more complete expression. 
 
\section{Discretized Position Eigenstates}

The derivation in the previous section was performed in the ``usual'' case 
where space is compactified on a circle of 
radius $r$, and the above expansion should be valid when $r$ is large compared
to the dispersion in the state $\Delta x^2$. By interchanging the roles of 
configuration space and momentum space one obtains a discrete spectrum for 
position space. We define an algebra as 
\begin{eqnarray}
\left [ G_p, W_p\right ]=W_p.
\label{comm}
\end{eqnarray}
for Hermition $G_p$ and unitary $W_p$.
The minimum uncertainty wave packet in momentum space,
\begin{eqnarray}
\psi_p(\theta_p)={1\over {\sqrt{I_0(2\beta_p)}}}\exp \left 
[\beta_p e^{i(\theta_p -\phi_p)}
+i\nu_p \theta_p \right ]\;,
\label{muwp}
\end{eqnarray}
is centered at $\phi_p$ and in the limit $\beta_p\to \infty$ one gets the
Gaussian. It is important to note that $\psi_p$ is the momentum space
wave function rather than the position space wave function $\psi$ considered
earlier.
The expectation values are the same as before
\begin{eqnarray}
\langle W_p \rangle &=&\rho_p e^{i\phi_p}\;, \nonumber \\
\langle G_p \rangle &=&\nu_p +\alpha_p +\beta_p \rho_p\;,
\end{eqnarray}
with 
\begin{eqnarray}
\rho_p={{I_1(2\beta_p)}\over {I_0(2\beta_p)}}\;.
\end{eqnarray}
Since the roles of position and momentum have been interchanged, we
define the variables $p$, $\langle p \rangle$ and 
$\langle x \rangle$
as 
\begin{eqnarray}
p&=&r_p\theta_p\;,\nonumber \\
\langle p \rangle &=& r_p\phi_p\;, \\
\langle x \rangle &=& {1\over r_p}\langle G_p \rangle\;. \nonumber 
\end{eqnarray}
One obtains a discrete spectrum for $G_p$ with consecutive eigenvalues 
separated by $1/r_p$ so that position space is discretized.
Repeating the steps in the previous section with the roles of position and 
momentum interchanged, one arrives at an analogous expression
expansion which has the form of a GUP as in Eq.~(\ref{gup}),
\begin{eqnarray}
\Delta x^2 \Delta p^2 ={1\over 4}\left (1+{1\over {4r_p^2}}
\Delta p^2+{1\over {12r_p^2}}{{\langle p^4\rangle}\over 
{\Delta p^2}}
+\dots\right )
\;,
\label{gup2}
\end{eqnarray}
where $r_p$ is the compactification radius of momentum space.
If we want the discretization of configuration space to be at a certain scale,
say $L_P$, (which results when we require the parameter $\nu_p$ 
to be an integer) 
then $r_p$ is determined in terms of $L_P$.

\begin{figure}[t]
\centerline{
\mbox{\includegraphics[width=3.50in]{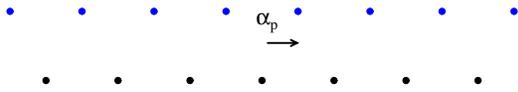}}
}
\caption{The parameter $\alpha_p$ characterizes the inequivalent 
discretizations of the position operator $G_p$. It takes values in the 
range $\alpha_p \in [0,1)$. The separation of eigenvalues of $G_p$ is given
in terms of the compactification radius as $1/r_p$ ($\hbar =1$).}
\label{discrete}
\end{figure}

A gauge parameter $\alpha_p$ interpolates between discretizations.
It takes on values in the range $\alpha_p \in [0,1)$.
The operator $G_p$ is represented by
\begin{eqnarray}
\langle \theta_p|G_p|\psi_p \rangle &=&\left [-i{\partial 
\over {\partial \theta_p}}
-i\kappa_p^*(\theta_p){{\partial \kappa_p(\theta_p)}\over {\partial \theta_p}}
+\alpha_p \right ]\psi_p\;,\nonumber \\
&\equiv& \left [-i{\partial 
\over {\partial \theta_p}}+A_p(\theta_p)\right ]\psi_p\;,
\label{Gpos}
\end{eqnarray}
so we can understand the parameter $\alpha_p$ as a quantity that interpolates
between the discrete eigenvalues of $G_p$ as shown 
in Fig.~\ref{discrete}. Choosing $\alpha_p$ to be an integer
represents a shift in the lattice and is physically equivalent to the 
case $\alpha_p=0$. The periodic function $\kappa_p(\theta_p)$ 
represents a gauge choice, and for simplicity it can be chosen to be equal 
to one in which case it disappears from Eq.~(\ref{Gpos}). 
It can be shown\cite{Tanimura:1993hf} that any periodic
function $A_p(\theta_p)=A_p(\theta_p+2\pi)$ is gauge equivalent to a constant 
function $\alpha_p$, and that two constant functions are gauge equivalent if 
they differ by an integer. This has an interesting interpretation for
a discretized space as we are considering here where it can be viewed as 
parameterizing the ways to populate equally spaced eigenvalues on the line.

\section{Expectation Values for the Minimum Uncertainty Wave Packets}

In this section we calculate expectation values for the minimal 
uncertainty wave packets that appear in the uncertainty relation in 
Eq.~(\ref{gup2}).
In particular we show how the usual minimum
uncertainty wave packet 
approaches the limit where the radius become very small. We also 
give examples of states which do not have any analog with respect to the 
usual Heisenberg uncertainty principle (because they are localized at 
discrete eigenvalues of the position operator).

\begin{figure}[b]
\centerline{
\mbox{\includegraphics[width=3.50in]{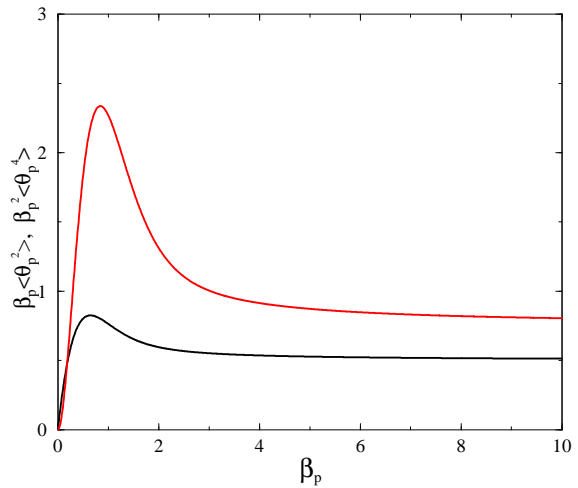}}
}
\caption{The quantities $\beta_p\langle \theta_p^2\rangle$ 
(lower curve) and $\beta_p^2\langle \theta_p^4\rangle$ (upper curve)
for the minimum uncertainty wave packet in 
Eq.~(\ref{muwp}) as a function of $\beta_p$. For large $\beta_p$ they approach 
the expected values of $1/2$ and $3/4$ respectively.}
\label{disp}
\end{figure}

The dispersion can be calculated for the minimum uncertainty wave packet in
Eq.~(\ref{muwp}). One finds 
\begin{eqnarray}
\langle \theta_p^2 \rangle = 
{\pi^2\over 3}+\sum _{n=1}^\infty {4\over n^2}(-1)^n
{{I_n(2\beta_p)}\over {I_0(2\beta_p)}}\;,
\end{eqnarray}
and 
\begin{eqnarray}
\langle \theta_p^4 \rangle = {\pi^4\over 5}+\sum _{n=1}^\infty \left [
{8\pi^2\over n^2}-{48\over n^4}\right ](-1)^n
{{I_n(2\beta_p)}\over {I_0(2\beta_p)}}
\;,
\end{eqnarray}
These quantities decrease from $\pi^2/3$ and $\pi^4/5$ to zero as $\beta_p$ 
goes from zero to infinity. This merely reflects the fact that the minimum
uncertainty wave packets
become more localized (squeezed) in the angular parameter $\theta_p$.
Multiplying by the appropriate power of $\beta_p$ results in functions that
asymptote to nonzero values for large $\beta_p$ as shown in Fig.~\ref{disp}.
It is clear that these two expectation values 
give comparable contributions to the 
right hand side of Eq.~(\ref{gup2}).

For the large radius limit one can calculate 
\begin{eqnarray}
\beta_p\langle \theta_p^2\rangle ={1\over 2}\;, \qquad
\beta_p^2\langle \theta_p^4\rangle ={3\over 4}\;,
\label{asymp}
\end{eqnarray}
by using the Gaussian form. These formulas receive ${\cal O}(1/\beta_p)$ 
corrections for large $\beta_p$. The approach to the 
asymptotic limits are shown 
in Fig.~\ref{disp}. From Eq.~(\ref{asymp}) one has 
$\langle p^4\rangle =3\langle p^2\rangle ^2 =3(\Delta p^2)^2$ 
for the minimum uncertainty wave packet
and the equality in Eq.~(\ref{gup2}) becomes 
\begin{eqnarray}
\Delta x^2 \Delta p^2 ={1\over 4}\left (1+{1\over {2r_p^2}}\Delta p^2
+\dots\right )
\;,
\label{gup3}
\end{eqnarray}
where the ellipsis refers to terms of order $1/r_p^4$ and higher.
For a more general state that does not necessarily saturate the uncertainty 
relation, one does not expect the ratio 
$\langle p^4\rangle /\langle p^2\rangle^2$ to be a fixed constant (for the 
minimum uncertainty wave packet it is 3).

The GUP in Eq.~(\ref{gup3}) describes the relationship between the 
dispersion in $x$ and $p$ in the limit where the discrete eigenvalues of 
the position operator $G_p$ are very finely spaced.
As $\beta_p$ is decreased from some large value,
the finite size of the circle allows the wave function
to attain some nonzero value for all values of the angular parameter $\theta_p$
(the wave function can
``see'' itself around the circle). When $\beta_p\to 0$, 
the Gaussian wave packet
has become the state with equal probability for each point on the circle. This
is the eigenstate of operator $G_p$ corresponding to position. 
Since $\theta_p$ is bounded by 
the periodicity of the circle, the right hand side 
of the uncertainty relation goes to zero. 
In our interpretation of the discrete space eigenvalue operator, the 
Gaussian which corresponds to the $\beta_p \to \infty$ case is smoothly 
interpolated to an eigenstate of the position operator as $\beta_p \to 0$.
During this interpolation the initial corrections to the right hand side 
of the uncertainty principle are positive as expected. Eventually when 
$\beta_p$ becomes small enough, the expansion in terms of $\Delta p$ is no 
longer adequate to approximate the right hand side of the uncertainty 
principle, Eq.~(\ref{unc}). When $\beta_p$ reaches zero, one obtains
an eigenstate of the position operator $G_p$:
\begin{eqnarray}
\psi_p(\theta_p)=\exp(i\nu_p \theta_p)\;,
\end{eqnarray}
and is 
characterized by a uniform distribution in (compactified) momentum space. 

\section{Summary and Conclusions}

We have derived the generalized uncertainty principle from a toy model of 
discretized space by considering quantum mechanics on a circle where the 
compacification involves the {\it momentum}. This model may be useful
in exploring how the ultraviolet limit is approached in more realistic 
models of discrete spacetime or models of quantum gravity with a 
fundamental or minimum length.

A GUP contains not only an infinite series
of higher powers of the momentum dispersion, but also typically involves
contributions from higher order quantities such as $\langle p^4 \rangle$.
In fact this quantity may be of the same order as the one usually 
encountered in the definition of the GUP. It is probably worthwhile to pursue
some models that fully capture this infinite series in addition to the more
common procedure of investigating the implications of the truncated GUP in 
terms of deformed algebra. From our perspective the defining algebra in 
Eq.~(\ref{comm}) leads to the uncertainty principle in Eq.~(\ref{unc0}), and
the truncated GUP arises as the appropriate formula in a certain physically 
interesting limit.

The large radius limit of the defining algebra yields a 
discretized position operator with finely-spaced eigenvalues.
It seems that the simple discretization of space implied by a compactified 
momentum is enough to obtain the leading order terms in the generalized 
uncertainty principle in a natural way. 
We find it interesting that a detailed incorporation
of gravity does not seem to be necessary to obtain the GUP. In the 
quantum mechanics on the circle the uncertainty relation is known for all
values of the momentum compactification radius, so physically what happens
to the variances of physical operators is known completely from one 
limit (an almost-Gaussian localized to a small region of the momentum circle) 
to the other (an eigenstate of the position operator). This may 
result in an improved understanding of the origin of the generalized 
uncertainty principle in theories of quantum gravity.

\section*{Acknowledgments}
This work was supported in part by the U.S.
Department of Energy under Grant No.~DE-FG02-91ER40661.


\begin{thebibliography}{xx}
\def\etal {{\it et al.}}

\bibitem{Veneziano}
G.~Veneziano, Europhys.\ Lett.\ {\bf 2}, 199 (1986). 

\bibitem{Gross:1987ar}
  D.~J.~Gross and P.~F.~Mende,
  Nucl.\ Phys.\ B {\bf 303}, 407 (1988).

\bibitem{Amati:1988tn}
  D.~Amati, M.~Ciafaloni and G.~Veneziano,
  Phys.\ Lett.\ B {\bf 216}, 41 (1989).

\bibitem{Konishi:1989wk}
  K.~Konishi, G.~Paffuti and P.~Provero,
  Phys.\ Lett.\ B {\bf 234}, 276 (1990).

\bibitem{Maggiore:1993rv}
  M.~Maggiore,
  Phys.\ Lett.\ B {\bf 304}, 65 (1993)
  [arXiv:hep-th/9301067].

\bibitem{Snyder:1946qz}
  H.~S.~Snyder,
  Phys.\ Rev.\  {\bf 71}, 38 (1947).

\bibitem{Silvert:1970kr}
  W.~Silvert,
  Phys.\ Rev.\ D {\bf 2}, 633 (1970);
  M.~Maggiore,
  Phys.\ Rev.\ D {\bf 49}, 5182 (1994)
  [arXiv:hep-th/9305163];
  M.~Maggiore,
  Phys.\ Lett.\ B {\bf 319}, 83 (1993)
  [arXiv:hep-th/9309034];
  M.~Maggiore,
IFUP-TH-27-94
{\it Lectures given at 30th Karpacz Winter School of Theoretical Physics: Quantum Groups: Formalism and Applications, Karpacz, Poland, 14-26 Feb
1994};
  L.~J.~Garay,
  Int.\ J.\ Mod.\ Phys.\ A {\bf 10}, 145 (1995)
  [arXiv:gr-qc/9403008];
  C.~Castro,
  Found.\ Phys.\ Lett.\  {\bf 10}, 273 (1997)
  [arXiv:hep-th/9512044];
  A.~Kempf, G.~Mangano and R.~B.~Mann,
  Phys.\ Rev.\ D {\bf 52}, 1108 (1995)
  [arXiv:hep-th/9412167];
  R.~Simon and N.~Mukunda,
  arXiv:quant-ph/9708037;
  F.~Scardigli,
  Phys.\ Lett.\ B {\bf 452}, 39 (1999)
  [arXiv:hep-th/9904025];
  S.~Capozziello, G.~Lambiase and G.~Scarpetta,
  Int.\ J.\ Theor.\ Phys.\  {\bf 39}, 15 (2000)
  [arXiv:gr-qc/9910017];
  L.~B.~Crowell,
  Found.\ Phys.\ Lett.\  {\bf 12}, 585 (1999);

\bibitem{Adler:2001vs}
  R.~J.~Adler, P.~Chen and D.~I.~Santiago,
  Gen.\ Rel.\ Grav.\  {\bf 33}, 2101 (2001)
  [arXiv:gr-qc/0106080];
  S.~Kalyana Rama,
  Phys.\ Lett.\ B {\bf 519}, 103 (2001)
  [arXiv:hep-th/0107255];
  A.~Camacho,
  Gen.\ Rel.\ Grav.\  {\bf 34}, 1839 (2002)
  [arXiv:gr-qc/0206006];
  F.~Nasseri,
  Phys.\ Lett.\ B {\bf 618}, 229 (2005)
  [arXiv:astro-ph/0208222];
  A.~Camacho,
  Rel.\ Grav.\ Cosmol.\  {\bf 1}, 89 (2004)
  [arXiv:gr-qc/0302096];
  A.~Camacho,
  Gen.\ Rel.\ Grav.\  {\bf 35}, 1153 (2003)
  [arXiv:gr-qc/0303061];
  P.~Chen,
  arXiv:astro-ph/0305025;
  P.~S.~Custodio and J.~E.~Horvath,
  Class.\ Quant.\ Grav.\  {\bf 20}, L197 (2003)
  [arXiv:gr-qc/0305022];
  F.~Scardigli and R.~Casadio,
  Class.\ Quant.\ Grav.\  {\bf 20}, 3915 (2003)
  [arXiv:hep-th/0307174];
  C.~Z.~Liu, X.~Li and Z.~Zhao,
  Int.\ J.\ Theor.\ Phys.\  {\bf 42}, 2081 (2003);
  M.~R.~Setare,
  Phys.\ Rev.\ D {\bf 70}, 087501 (2004)
  [arXiv:hep-th/0410044];
  B.~Bolen and M.~Cavaglia,
  Gen.\ Rel.\ Grav.\  {\bf 37}, 1255 (2005)
  [arXiv:gr-qc/0411086];
  C.~Li, L.~Xiang and Y.~G.~Shen,
  Int.\ J.\ Mod.\ Phys.\ D {\bf 13}, 1847 (2004);

\bibitem{Medved:2004yu}
  A.~J.~M.~Medved and E.~C.~Vagenas,
  Phys.\ Rev.\ D {\bf 70}, 124021 (2004)
  [arXiv:hep-th/0411022];
  M.~Maziashvili,
  Phys.\ Lett.\ B {\bf 635}, 232 (2006)
  [arXiv:gr-qc/0511054];
  M.~T.~Makhviladze, M.~A.~Maziashvili and D.~V.~Nozadze,
  arXiv:gr-qc/0512044.

\bibitem{Nozari:2005it}
  K.~Nozari and T.~Azizi,
  Int.\ J.\ Quant.\ Inf.\  {\bf 3}, 623 (2005)
  [arXiv:gr-qc/0504090];
  M.~R.~Setare,
  Int.\ J.\ Mod.\ Phys.\ A {\bf 21}, 1325 (2006)
  [arXiv:hep-th/0504179];
  F.~Scardigli and R.~Casadio,
  Braz.\ J.\ Phys.\  {\bf 35}, 470 (2005);
  F.~Nasseri,
  Phys.\ Lett.\ B {\bf 632}, 151 (2006)
  [arXiv:hep-th/0510184];
  T.~Matsuo and Y.~Shibusa,
  Mod.\ Phys.\ Lett.\ A {\bf 21}, 1285 (2006)
  [arXiv:hep-th/0511031];
  K.~Nozari and B.~Fazlpour,
  arXiv:gr-qc/0601092;
  K.~Nozari,
  Int.\ J.\ Theor.\ Phys.\  {\bf 44}, 1325 (2005);
  W.~Kim, Y.~W.~Kim and Y.~J.~Park,
  arXiv:gr-qc/0605084;
  F.~Scardigli and R.~Casadio,
{\it Prepared for 5th International Workshop on New Worlds in Astroparticle Physics, Faro, Portugal, 8-10 Jan 2005};
  Y.~Ko, S.~Lee and S.~Nam,
  arXiv:hep-th/0608016;
  Z.~Hai-Xia, L.~Huai-Fan, H.~Shuang-Qi and Z.~Ren,
  arXiv:gr-qc/0608023.

\bibitem{Chang:2001kn}
  L.~N.~Chang, D.~Minic, N.~Okamura and T.~Takeuchi,
  Phys.\ Rev.\ D {\bf 65}, 125027 (2002)
  [arXiv:hep-th/0111181];
  L.~N.~Chang, D.~Minic, N.~Okamura and T.~Takeuchi,
  Phys.\ Rev.\ D {\bf 65}, 125028 (2002)
  [arXiv:hep-th/0201017];
  S.~Benczik, L.~N.~Chang, D.~Minic, N.~Okamura, S.~Rayyan and T.~Takeuchi,
  Phys.\ Rev.\ D {\bf 66}, 026003 (2002)
  [arXiv:hep-th/0204049];
  S.~Benczik, L.~N.~Chang, D.~Minic, N.~Okamura, S.~Rayyan and T.~Takeuchi,
  arXiv:hep-th/0209119;
  R.~Akhoury and Y.~P.~Yao,
  Phys.\ Lett.\ B {\bf 572}, 37 (2003)
  [arXiv:hep-ph/0302108];
  S.~Hossenfelder, M.~Bleicher, S.~Hofmann, J.~Ruppert, S.~Scherer and H.~Stoecker,
  Phys.\ Lett.\ B {\bf 575}, 85 (2003)
  [arXiv:hep-th/0305262];
  S.~Hossenfelder,
  Class.\ Quant.\ Grav.\  {\bf 23}, 1815 (2006)
  [arXiv:hep-th/0510245];
  S.~Hossenfelder,
  Phys.\ Rev.\ D {\bf 73}, 105013 (2006)
  [arXiv:hep-th/0603032].

\bibitem{Kempf:2000ac}
  A.~Kempf,
  Phys.\ Rev.\ D {\bf 63}, 083514 (2001)
  [arXiv:astro-ph/0009209];
  A.~Ashoorioon, A.~Kempf and R.~B.~Mann,
  Phys.\ Rev.\ D {\bf 71}, 023503 (2005)
  [arXiv:astro-ph/0410139].



\bibitem{London}
F.~London, Zeitschr.\ f.\ Physik {\bf 37}, 915 (1926);
Zeitschr.\ f.\ Physik {\bf 40}, 193 (1927).

\bibitem{Jordan}
P.~Jordan, Zeitschr.\ f.\ Physik {\bf 40}, 809 (1927);
Zeitschr.\ f.\ Physik {\bf 44}, 1 (1927).

\bibitem{Kastrup:2003fs}
  H.~A.~Kastrup,
  Fortsch.\ Phys.\  {\bf 51}, 975 (2003)
  [arXiv:quant-ph/0307069].

\bibitem{Judge}
D.~Judge, Nuovo\ Cimento\ {\bf 31}, 332 (1964).

\bibitem{Bouten}
M.~Bouten, N.~Maene and P.~Van~Leuven, Nuovo\ Cimento\ {\bf 37}, 1119 (1965).

\bibitem{Chisolm}
E.~D.~Chisolm, 
Am.\ J.\ Phys.\ {\bf 69}, 368 (2001).

\bibitem{Susskind} 
L.~Susskind and J.~Glogower, Physics {\bf 1}, 49 (1964).

\bibitem{Carruthers:1968my}
  P.~Carruthers and M.~M.~Nieto,
  Rev.\ Mod.\ Phys.\  {\bf 40}, 411 (1968).

\bibitem{Jackiw}
R.~Jackiw,
J.\ Math.\ Phys.\ {\bf 9}, 339 (1968).

\bibitem{Levy-Leblond:1976uj}
  J.~M.~Levy-Leblond,
  Annals Phys.\  {\bf 101}, 319 (1976).

\bibitem{Newton:1979ut}
  R.~G.~Newton,
  Annals Phys.\  {\bf 124}, 327 (1980).

\bibitem{Ohnuki:1991uv}
  Y.~Ohnuki and S.~Kitakado,
  Mod.\ Phys.\ Lett.\ A {\bf 7}, 2477 (1992);
  Y.~Ohnuki and S.~Kitakado,
  J.\ Math.\ Phys.\  {\bf 34}, 2827 (1993);
  Y.~Ohnuki and S.~Kitakado,
  Mod.\ Phys.\ Lett.\ A {\bf 9}, 143 (1994).

\bibitem{Tanimura:1993hf}
  S.~Tanimura,
  Prog.\ Theor.\ Phys.\  {\bf 90}, 271 (1993)
  [arXiv:hep-th/9306098].

\bibitem{Barut:1970qf}
  A.~O.~Barut and L.~Girardello,
  Commun.\ Math.\ Phys.\  {\bf 21}, 41 (1971).

\bibitem{Bluhm:1995nr}
  R.~Bluhm, V.~A.~Kostelecky and B.~Tudose,
  Phys.\ Rev.\ A {\bf 52}, 2234 (1995)
  [arXiv:quant-ph/9509010];
  R.~Bluhm, V.~A.~Kostelecky and B.~Tudose,
  Phys.\ Rev.\ A {\bf 53}, 937 (1996)
  [arXiv:quant-ph/9510023].
  V.~A.~Kostelecky and B.~Tudose,
  Phys.\ Rev.\ A {\bf 53}, 1978 (1996)
  [arXiv:quant-ph/9512030].

\bibitem{Trifonov:2000su}
  D.~A.~Trifonov,
  J.\ Opt.\ Soc.\ Am.\ A {\bf 17}, 2486 (2000)
  [arXiv:quant-ph/0012072].

\bibitem{Kastrup:2001hx}
  H.~A.~Kastrup,
  arXiv:quant-ph/0109013;
  H.~A.~Kastrup,
  Phys.\ Rev.\ A {\bf 73}, 052104 (2006)
  [arXiv:quant-ph/0510234].

\bibitem{Gonzalez:1998kj}
  J.~A.~Gonzalez and M.~A.~del Olmo,
  J.\ Phys.\ A {\bf 31}, 8841 (1998)
  [arXiv:quant-ph/9809020].

\bibitem{Kowalski:1998hx}
  K.~Kowalski, J.~Rembielinski and L.~C.~Papaloucas,
  J.\ Phys.\ A {\bf 29}, 4149 (1996)
  [arXiv:quant-ph/9801029].

\bibitem{Roy:1982nu}
  S.~M.~Roy and V.~Singh,
  Phys.\ Rev.\ D {\bf 25}, 3413 (1982).

\bibitem{Santhanam:1976zs}
  T.~S.~Santhanam,
  Phys.\ Lett.\ A {\bf 56}, 345 (1976);
  R.~Jagannathan, T.~S.~Santhanam and R.~Vasudevan,
  Int.\ J.\ Theor.\ Phys.\  {\bf 20}, 755 (1981);
  R.~Jagannathan and T.~S.~Santhanam,
  Int.\ J.\ Theor.\ Phys.\  {\bf 21}, 351 (1982).

\bibitem{Isham}
C.~J.~Isham, Topology and global aspects of quantum theory. In {\it Relativity,
Groups and Topology II}, Eds. B.~S.~DeWitt and R.~Stora, (1984).

\end{thebibliography}
\end{document}